\documentclass[aps,prl,preprint,unsortedaddress,showkeys,nofootinbib]{revtex4-1}

\usepackage{amssymb,amsmath,amsfonts}

\usepackage{graphics}
\usepackage{graphicx}
\usepackage{epstopdf}
\usepackage[pdftex]{hyperref}
\usepackage{longtable}

\usepackage{bm}

\usepackage{color}
\usepackage{setspace}
\usepackage{multirow}

\usepackage[ruled,vlined,linesnumbered]{algorithm2e}

\usepackage{color, colortbl}
\definecolor{Yellow}{rgb}{1,1,0}
\definecolor{Grey}{rgb}{.87,.87,.87}
\definecolor{Purple}{rgb}{.8,.0,1.0}
\definecolor{Crimson}{rgb}{.86,.08,.23}

\begin{document}

\title{Complex network representation through multi-dimensional \\ node projection}

\author{Stanislav Sobolevsky}
\affiliation{Urban Complexity Group, Center For Urban Science+Progress, \\
New York University\\
sobolevsky@nyu.edu, www.ucomp.net}


\begin{abstract}
\begin{it}
Complex network topology might get pretty complicated challenging many network analysis objectives, such as community detection for example. This however makes common emergent network phenomena such as scale-free topology or small-world property even more intriguing. In the present proof-of-concept paper we propose a simple model of network representation inspired by a signal transmission physical analogy, which is apparently capable of reproducing both of the above phenomena. The model appears to be general enough to represent and/or approximate arbitrary complex networks. We propose an approach constructing such a representation by projecting each node into a multi-dimensional space of signal spectrum vectors, where network topology is induced by their overlaps. As one of the implications this enables reducing community detection in complex networks to a straightforward clustering over the projection space, for which multiple efficient approaches are available. We believe such a network representation could turn out to be a useful tool for multiple network analysis objectives. \end{it}
\end{abstract}


\maketitle

\section{Introduction}
Complex networks start to penetrate multiple fields of science such as physics, biology, economics, social sciences, urban planning as they describe features of the increasingly interconnected world, such as physical and digital infrastructure, biological interactions, economic transactions as well as human mobility and communications. This makes approaches revealing the structure of the complex networks relevant to all the domains above. 

Among such approaches one of the most common ones is community detection \cite{fortunato2010}. Community detection saw a wide range of applications in social science \cite{plantie2013survey}, biology\cite{Guimera2005FunctionalCartography}, economics \cite{PiccardiWorldTradeWeb}, studies of human mobility and interactions with applications, for example, to regional delineation \cite{Ratti2010GB, blondel2010regions, Sobolevsky2013delineating, amini2014impact, hawelka2014geo, kang2013exploring, sobolevsky2014money, belyi2017global, grauwin2017identifying, sobolevsky2017inferring, sobolevsky2018twitter}.

However community structure alone is far from being able to provide a comprehensive characteristic of the network topology. The present paper will further contribute towards this objective by providing a network representation approach potentially able to capture more complicated phenomena, while also applicable to community detection in particular. We start with a random network model inspired by signal transmission which apparently turns out to be capable of reproducing most common network phenomena such as scale-free topology \cite{Barabasi2009scale} and small-world property \cite{watts1998collective}. Further we show that this model can be used to represent or approximate arbitrary unweighted and weighted networks. Finally we demonstrate how such approximation could contribute to community detection by replacing complicated heuristics by straightforward clustering algorithms in a new network projection space.

\section{The signal spectrum network model.}
Employ a following physical analogy: let each network node broadcast and receive signals of a certain discrete spectrum of frequencies and connect those pairs of nodes which receive at least one type of signals from each other. Mathematically, such nodes could be characterized by binary vectors, encoding each transmitted/received signal frequency within the spectrum as 1 and missed frequencies as 0, while network edges are put between those pairs which share at least one unit in their binary vector representations (projections). Call such a model a binary signal spectrum model (BSSM).

Now generate node representations of a BSSM as vectors of independent Bernoulli random variables with the same fixed probability $p$ of having a unit in each spot and connect the nodes according to the above rule. Loop edges are ignored. As we show below, under certain conditions (basically having sufficient density of units to ensure connectivity throughout the network) the resulting BSSM network possesses clear scale-free topology as well as the small-world property.

Specifically after 10 simulations of a 1000 node network with 100-dimensional node signal spectrum (100 types of signals) and a probability $p=2\%$ for each node to have each type of signal transmitted and received, we get node degrees ranging from 0 to approximately 300 with the largest principle connected component covering almost the entire network (component sizes range around 950 in each simulation). In each simulation the top 50 node degree distributions approximately follow a power law $d\sim r^q$ where $r$ is the node rank, the exponent $q=-0.129\pm 0.037$ (average and standard deviations of 10 simulations). The average distribution for 10 simulations almost perfectly resembles a line on a log-log scale as illustrated on the figure 1 below. This way one can claim the constructed model to be a scale-free network. Worth mentioning that the entire degree distribution has a heavy tail and is more likely to follow a log-normal distribution (which is also pretty common for scale-free networks which often follow power law degree distribution only for the top nodes).

Within the largest connected components average distance is $2.182\pm 0.026$, while the maximal distance is $3.273\pm 1.191$ accordingly. So on average each pair of nodes is reachable from each other in 2-3 steps, at most in 3-5 steps, perfectly resembling the small world phenomena.

\begin{figure}[h]
\centering
\includegraphics[width=0.6\textwidth]{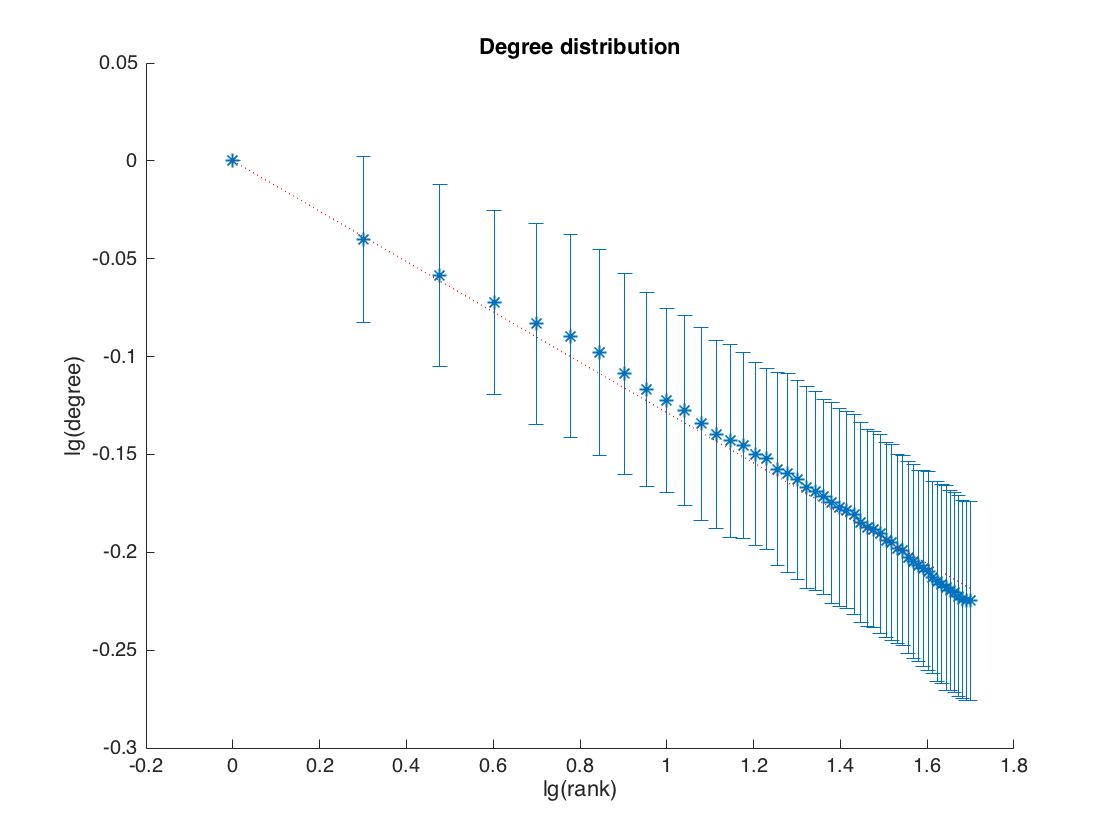}
\caption{\label{fig::SampleNet}Average degree distribution (log-degree averages with standard deviation error bars for top 50 nodes for all the networks generated) for 10 sample networks with 1000 nodes, 100-dimensional node representation, $p=0.02$, approximated with a power law $log(d)\sim q \cdot log(r)+log(\max d)$. Degrees distributions in each simulation approximately follow power laws with exponents $q=-0.129\pm 0.037$ (average/standard deviation), average distance and maximal distance within maximal connected component are $2.182\pm 0.026$ and $3.273\pm 1.191$ accordingly.} 
\end{figure}

\section{Representing real-world networks}

Not only the above model could serve as an interesting artificial random network example resembling key properties of the real-world networks, but more importantly it could be leveraged to represent arbitrary complex networks. In order to do so one needs to propose a signal spectrum transmitted and received by each node, such that edges connecting the nodes in accordance to the signal spectrum model resemble the actual edges of the original network.

\subsection{Representing unweighed networks with BSSM.}

Start with a binary (unweighed) undirected case and use the above binary signal spectrum model. First show that the model is capable to represent any such network in principle. 

{\bf Theorem 1.} {\it Every unweighted undirected network of size $n$ could be exactly represented through a binary signal spectrum model of dimension $m$, where $m$ is the minimal number of (possible overlapping cliques) needed to cover the network.}

Edges of each graph could be covered by a union of (possibly overlapping) fully connected cliques \cite{roberts1985applications}. The number $m$ of such cliques is known to be no higher than the number of edges in the network (as one could consider each edge to be a separate clique) or $n^2/4$ \cite{erdos1966representation} whichever is lower. Then assigning a dedicated signal type (binary digit) to represent each clique we encode each node with a binary sequence having units in the positions representing the cliques the node belongs to. This way the pair of nodes appear to be connected if and only if it belong to at least one of the cliques, which is equivalent to being connected with an edge of the original network. This way the proof of theorem 1 is complete.

For example the well-known Zachary Karate club network \cite{zachary1977ifm} of size $n=34$ could be represented with as low as $35$ overlapping cliques and this way could be represented with a 35-dimensional BSSM.

\subsection{Representing weighed networks with WSSM.}

For more efficient and practical representations, including representations of the arbitrary weighted (including negative weighs) undirected or directed networks, a more general model could be introduced. Above BSSM model represented each node with its membership in the cliques covering the network. Now represent each node $j$ with a vector $w_j=(w_j^1,w_j^2,...,w_j^m)$ of arbitrary real numbers, representing strength of the signals transmitted/received or node membership in each clique with respect to its strength, allowing arbitrary positive or negative strength (attraction or repulsion to each clique). For directed networks assign each node two separate vectors $w_j^{out}$ and $w_j^{in}$ - one for outgoing and one for incoming edge (signal) representation (this way the spectra of transmitted and received signals could be different). Call such as model a weighted directed signal spectrum model (WDSSM). In the undirected case let $w_j^{out}=w_j^{in}=w_j$ and call this a weighted undirected signal spectrum model (WUSSM or simply WSSM).

This way a network edge $e_{i,j}$ between the nodes $i,j$ could be represented or approximated as
\begin{equation}
e_{i,j}=\sum_{d=1}^m w_i^{out, d}w_j^{in, d}.
\label{refWSSM}
\end{equation}

Approximation (\ref{refWSSM}) for symmetric networks including loop edges appears to be straightforward: it could be shown that the best approximation is provided by the $w^j$ being the leading eigenvectors of the network's adjacency matrix. Situation becomes more complicated in the directed case and especially in case when loop edges are excluded from consideration and fitting then is not essential. Nevertheless, eigenvector decomposition still turns out to be helpful for designing a quickly converging heuristic algorithm for the case of omitted loop edges.

Although primarily designed to represent weighted networks, WSSM might be helpful even for unweighted case providing a more compact representation with lower $m$ compared to BSSM. In this case it is might not be necessary to fit the binary weights $e_{i,j}$ precisely, but approximate representation could also be sufficient to represent the network subject to rounding or applying a certain classification threshold: 
$$
e_{i,j}=1 \Leftrightarrow \sum_{d=1}^m w_i^{out, d}w_j^{in, d}\geq \theta,
$$  
where $\theta=0.5$ or other. Then for unweighted networks it is actually sufficient to use integer weights $w_i^{in/out, d}$ subject to appropriate integer threshold $\theta$. Call the above representation wighted integer approximation signal spectrum model $WIASSM$.

For example, Zachary's Karate Club \cite{zachary1977ifm} network could be exactly represented with a 14-dimensional $WIASSM$ (undirected). Our spectral algorithm based on iterative optimization of each $d-th$ layer of weights $w_j^d$ done simultaneously for for all $j$ (leveraging analytic matrix approximation utilizing eigenvalue decomposition) was able to find such a representation for $d=14$ but not for $d=13$, however since the algorithm appears to be a heuristic, a proof of that $d<14$ could not suffice remains an open question.  

\section{Application to community detection}
Exact network representation or approximation with WSSM could be useful for a variety of applications including hub detection (nodes with a broad signal spectrum), constructing novel spectrum-based centrality metrics as well as for community detection. So far one of the most common approaches for community detection is modularity optimization \cite{newman2004,newman2006}. 
Modularity of the partition $c(j)$ (a mapping assigning a certain community number $c$ to each node $j$) can be defined as 
\begin{equation}
Q=\sum_{i,j, c(i)=c(j)}q_{i,j},
\label{modularity}
\end{equation}
where the quantities $q_{i,j}$ for each edge $i,j$ (call them modularity scores of edges) are defined as
$$
q_{i,j}=\frac{e_{i,j}}{T}-\frac{k^{out}(i)k^{in}(j)}{T^2},
$$
where $k^{out}(i)=\sum_j e_{i,j}$, $k^{in}(j)=\sum_i e_{i,j}$, $T=\sum_i k^{out}(i)=\sum_j k^{in}(j)=\sum_{i,j}e_{i,j}$.
If the network is undirected then the edge modularity scores $q$ are symmetrical: $q_{i,j}=q_{i,j}$. However even for the directed case, the modularity scores could be effectively symmetrized assigning $q_{i,j}:=(q_{i,j}+q_{j,i})/2$ without any impact on the total score $Q$.

Since $WSSM$ allows to represent networks with both - positive and negative - edge weights, it could be directly applied to approximating modularity scores $q_{i,j}$ rather than the original edge weights $e_{i,j}$. The benefit of such an approximation is being able to replace cumulative in-community maximization of the arbitrary modularity scores by a clustering based on the cumulative vectorized distance maximization over a certain $m$-dimensional projection vector space. Modularity $Q$ of a certain partition $c_j$ over the WSSM-represented or approximated network could be represented or approximated as
$$
Q=\sum_{i,j, c_i=c_j}q_{i,j}=\sum_c\sum_{i,j, c_i=c_j=c}\sum_d w_i^{out, d}w_j^{in, d}=\sum_{c,d}\left(\sum_{i, c_i=c} w_i^{out, d}\right)\left(\sum_{j, c_j=c} w_j^{in, d}\right)=
$$$$
=\sum_{c} w_c^{out}\cdot w_c^{in},
$$
where $w_c^{in,out}$ are the aggregated incoming and outgoing signal strength vectors of the community $c$. The accuracy of representation or approximation of the loop edges is not relevant to the modularity maximization objective. In case of undirected networks 
$$
Q=\sum_{c} \|w_c\|^2.
$$
where $\|\cdot\|$ is the Euclidian norm of the vector. In case $d=1$ maximizing $Q$ becomes trivial~- one should take all $j$ with positive $w_j^1$ as the first community and the rest as the second community in order to produce a modularity-optimal partitioning. 

For the best single-dimensional WSSM approximation of Zachary's Karate Club \cite{zachary1977ifm} the trivial clustering above provides a bi-partition with the modularity score of 0.3715 (in terms of the original network scores) being pretty close to the best known bi-partitioning score of 0.3718 provided by \cite{combo} with a restriction on the resulting number of communities.

For $d>1$ the problem is more complicated but could be addressed through a clustering approach, similar to k-means clustering: starting with an arbitrary cluster assignments, iterate the procedure of computing cluster representations $w_c$ and re-assigning the nodes $j$ to maximize dot products $w_c \cdot w_j$ until no further adjustment of clusters is possible.

For example, for the Zachary's Karate Club \cite{zachary1977ifm} network this clustering approach over a 7 and higher-dimensional WSSM approximations of the modularity matrix allows to obtain a partitioning into 4 communities with a modularity score (in terms of the original network) of 0.4198, known to be the best possible modularity score (this is the best known partitioning score produced by \cite{combo} and it could be proven that a better score is not achievable \cite{sobolevsky2017optimality}). Worth mentioning that the current proof-of-concept implementation of the clustering approach is quite sensitive to the initial random cluster assignment and is susceptible to local extrema, so further improvement could be useful for insuring performance stability, which might be especially important for practical applications to larger networks.

\section{Conclusions}
This proof-of-concept paper starts from presenting a binary signal spectrum network model inspired by a physical analogy of signal transmission. Such a model being randomly generated is found capable of reproducing most prominent phenomena of the real-world networks - scale-free topology and small world property. The model is then generalized to a weighted signal spectrum model to represent or approximate arbitrary unweighted and weighted networks, projecting each node to a vector of transmitted/received signal strengths within a given signal spectrum. The network representation heuristic algorithm is proposed and illustrated on the example of the famous Zachary's Karate Club network, constructing its exact projection.  

We believe such a representation could have broad applications to pattern detection in complex networks. For example, once efficient network projection is constructed, community detection could be replaced by clustering the node projection vectors. E.g. for Zachary's Karate Club network, such a clustering over a 7-dimensional approximate projection space allows to find the optimal partitioning with the best possible modularity score.

\bibliographystyle{apsrev}

\end{document}